\documentclass[prl,twocolumn]{revtex4}

\usepackage{graphicx}
\usepackage{amsmath}
\begin{document}
\title{Resolving the HBT Puzzle in Relativistic Heavy Ion Collisions}
\author{Scott Pratt}
\affiliation{Department of Physics and Astronomy,
Michigan State University\\
East Lansing, Michigan 48824}
\date{\today}

\begin{abstract}
Two particle correlation data from the Relativistic Heavy Ion Collider have provided detailed femtoscopic information describing the space-time structure of the emission of pions. This data had avoided description with hydrodynamic-based approaches, in contrast to the success of hydrodynamics in reproducing other classes of observables. This failure has inspired the term ``HBT puzzle'', where HBT refers to femtoscopic studies which were originally based on Hanbury-Brown Twiss interferometry. Here, the puzzle is shown to originate not from a single shortcoming of hydrodynamic models, but the combination of several effects: mainly including pre-thermalized acceleration, using a stiffer equation of state, and adding viscous corrections.
\end{abstract}

\pacs{25.75.Gz,25.75.Ld}

\maketitle

Experiments at the Relativistic Heavy Ion Collider (RHIC) have revealed a new state of matter, the strongly interacting quark gluon plasma, which appears to have both perhaps the lowest ratio of viscosity to entropy of any measured substance \cite{Adcox:2004mh,Adams:2005dq}. This conclusions is based on comparisons of hydrodynamic models with experimental spectra and large angle correlations that reveal strong radial and elliptic collective flow, i.e., flow relative to the original beam axis. Whereas spectra and large angle correlations are consistent with ideal hydrodynamics \cite{Heinz:2002un,Teaney:2001av}, hydrodynamic models have poorly reproduced correlations at small-relative momentum \cite{Soff:2000eh,Teaney:2001av,Hirano:2002hv}. These correlations are related to the spatial and temporal properties of pion emission \cite{Lisa:2005dd}, and are often referred to a HBT (Hanbury-Brown Twiss) measurements after similar measurements with light \cite{HanburyBrown:1956pf}. It appears that hydrodynamic models underestimate the explosiveness of the collision, or equivalently overestimate the duration of the emission process. In contrast, some microscopic approaches have been more successful, though not completely satisfactory, in reproducing the data \cite{Petersen:2008gy,Li:2008ge,Humanic:2006sk,Lin:2002gc}. Unlike the hydrodynamic models, which employed first order phase transitions, the effective equations of state for the microscopic approaches are extremely stiff, leading to more explosive collisions. In short, the HBT puzzle involves finding whether one can reproduce femtoscopic observations with hydrodynamic models without using an equation of state that is inconsistent with lattice QCD calculations. In this study, we show this can be accomplished if three improvements are incorporated into hydrodynamic models: accounting for the buildup of collective flow in the first instants of the collision before thermalization is attained, using a stiffer equation of state, and including viscosity. The hydrodynamic model used here is outlined in detail in Ref. \cite{Pratt:2008sz}.

The Koonin equation \cite{Koonin:1977fh} relates the experimentally measured correlation function to the outgoing phase space density,
\begin{equation}\label{eq:koonin}
C({\bf P},{\bf q})=\int d^3r~S({\bf P},r)\left|\phi({\bf q},{\bf r})\right|^2.
\end{equation}
Here, $P$ and $q$ are the total and relative momentum. The source function $S({\bf P},{\bf r})$ describes the probability for two particles with identical momenta ${\bf k}={\bf P}/2$, to be separated by ${\bf r}$ in their asymptotic state if the relative interaction between the particles were to be ignored. Any dynamical model, whether based on hydrodynamics or microscopic degrees of freedom, provides a list of positions and times from which particles of specific ${\bf k}$ are emitted, and can then be used to generate the source function.

With some effort, the source function can be imaged from the correlation function, but for the purposes of this paper we will only consider parameters extracted by fitting to correlations that arise from a Gaussian source,
\begin{equation}
S({\bf P},{\bf r})\sim \exp\left\{-\frac{x^2}{4R_{\rm out}^2}-\frac{y^2}{4R_{\rm side}^2}
-\frac{z^2}{4R_{\rm long}^2}\right\}.
\end{equation}
The parameter $R_{\rm long}$ describes the longitudinal size along the beam axis, $R_{\rm side}$ describes the extent along the sideward dimension perpendicular to both the beam axis and ${\bf P}$. The outward dimension, $R_{\rm out}$ describes the size along the axis along the direction of ${\bf P}$, assuming one has boosted along the beam axis to a frame where ${\bf P}$ points perpendicular to the $x$ axis. Each of these dimensions can be a function of the transverse momentum $k_t=P_x/2$ or the longitudinal rapidity of the pair. For central collisions, which is what will be described here, the dimensions are independent of the azimuthal angle of ${\bf P}$. Furthermore, we will only consider central rapidity and only consider the $k_t$ dependence of the Gaussian dimensions. 

\begin{figure}
\centerline{\includegraphics[width=0.46\textwidth]{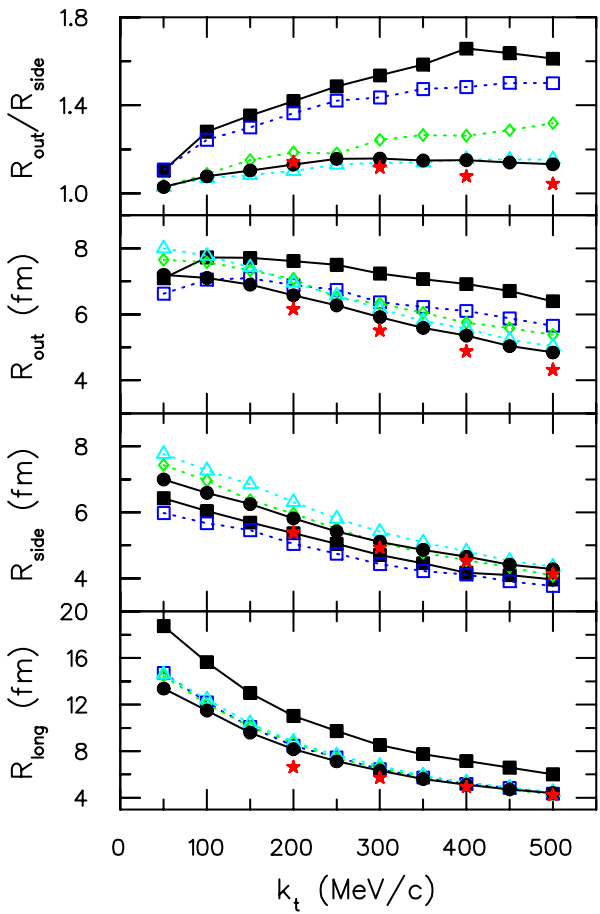}}
\caption{\label{fig:Rosl_everything}(color online)
Gaussian radii reflecting spatial sizes of outgoing phase space distributions in three directions: $R_{\rm out}$, $R_{\rm side}$ and $R_{\rm long}$. Data from the STAR collaboration (red stars) are poorly fit by a model with a first-order phase transition, no pre-thermal flow, and no viscosity (solid black squares). Correcting for all those deficiencies, and using a more appropriate treatment of the relative wave function in Eq. (\ref{eq:koonin}) brings calculations close to the data (filled black circles). The sequential effects of including prethermal acceleration (open blue squares), using a more realistic equation of state (open green diamonds), and adding viscosity (open cyan triangles) all make substantial improvements to fitting the data. An improved relative wave function yielded modest improvements (compare open cyan triangles to filled black circles).}
\end{figure}

Figure \ref{fig:Rosl_everything} shows experimentally determined radius parameters from $100 A$ GeV Au on $100 A$ GeV Au collisions at RHIC. For comparison, source sizes were generated from a hydrodynamic model coupled to a cascade code. The cascade microscopically simulates the final stages of the collision and breakup where local kinetic equilibrium is lost and hydrodynamics is unjustified. The times and positions of last collisions for particles of a specific ${\bf k}$ were used to calculate the source function, from which correlation functions were generated via Eq. (\ref{eq:koonin}). These were then fit to correlations from Gaussian sources to extract radii, which are also displayed in Fig. \ref{fig:Rosl_everything}.

As a benchmark, the first calculation (filled squares in Fig. \ref{fig:Rosl_everything}) was parameterized similarly to previous hydrodynamic calculations, and failed in a similar manner. Transverse expansion was delayed until 1 fm/$c$ after the initial collision. A strong first-order phase transition, which is inconsistent with lattice gauge theory, was employed, and the viscosities were set to zero. Additionally, an over-simplified relative wave function, neglecting Coulomb and strong interactions between the pions, was used to generate correlation functions. Since the source functions are not truly Gaussian, this can lead to different Gaussian radii. This benchmark calculation overstates the $R_{\rm out}/R_{\rm side}$ ratio by $\sim 40\%$ and overstates $R_{\rm long}$ by $\sim 25\%$.

The second calculation (open squares in Fig. \ref{fig:Rosl_everything}) accounts for prethermal acceleration by   beginning the expansion 0.1 fm/$c$ after the initial collision, roughly the amount of time required for the Lorentz contracted nuclei to traverse one another. The importance of pre-thermalized acceleration has been emphasized in several studies during the last few years \cite{Gyulassy:2007zz,Li:2008ge,Broniowski:2008vp}. As was shown in Ref. \cite{Vredevoogd:2008id}, flow during the first 1 fm/$c$ is approximately universal for any system with a traceless energy tensor, including partonic and field based pictures, independent of thermalization. Since the transverse expansion starts earlier, the longitudinal size is smaller at breakup, more in line with data. The $R_{\rm out}/R_{\rm side}$ ratios also drop, moving modestly toward the data.

The second improvement to be considered is to use a stiffer equation of state. Early studies used an equation of state with a first order phase transition with a large latent heat \cite{Soff:2000eh,Teaney:2001av,Hirano:2002hv}. Such soft equations of state have constant temperature and pressure for energy densities between $\epsilon_h$ and $\epsilon_h+L$, where $\epsilon_h$ is the maximum density of the hadronic phase. Here, $\epsilon_h$ corresponds to  a hadronic gas with a temperature of $T_c=170$ MeV, and $L$ is the latent heat. In contrast, lattice QCD now suggests a crossover transition where the pressure rises continuously with energy density. There indeed exists a soft region, but the speed of sound, $c_s^2=dP/d\epsilon$, never falls below 0.1 and the width of the soft region is somewhat lower than the latent heat $L$ assumed in the previous studies. The benchmark calculation, displayed in the upper panel, assumed a first order transition with a latent heat $L=1.6$ GeV/fm$^3$ with a lower bound to the mixed phase at $\epsilon_h\approx 500$ MeV/fm$^3$. This is not only inconsistent with lattice calculations, but is also inconsistent with femtoscopic analyses of data at lower energies. For heavy ion collisions at the upper AGS and for the lower SPS beam energies, maximum energy densities were in the neighborhood of $\epsilon_h+L$. For a first order phase transition the pressure $P$ stays fixed throughout the mixed phase, and these conditions would have minimal values of $P/\epsilon$ with minimal explosivity resulting in perhaps dramatically large lifetimes, well exceeding 20 fm/$c$. The long duration of the emission would lead to extended values of the outward dimensions of the phase-space cloud \cite{Pratt:1986cc,Rischke:1996em}. This was not observed. The third calculation (open diamonds in Fig. \ref{fig:Rosl_everything}) assumed a soft region of half the width in energy density, and with a speed of sound of $c_s^2=0.1$, rather than zero for a first-order transition. Once above the soft region, both calculations assumed a stiffening with the speed of sound $c_s^2=0.3$. One can consult Ref. \cite{hama} for a more sophisticated attempt at parameterizing lattice equations of state. As expected, the stiffer equation of state leads to more explosive collisions. A more sudden emission results in phase space clouds, $f(k_t,{\bf r},t\rightarrow\infty)$, that are less extended along the outward dimension, which lowers the the $R_{\rm out}/R_{\rm side}$ ratio. As can be seen in Fig. \ref{fig:Rosl_everything} the ratio moves significantly toward the data with the stiffer equation of state.

Shear viscosity is also known to increase the explosiveness of the collision \cite{Pratt:2008sc,Romatschke:2007mq}. This can be understood by considering viscous corrections to the stress-energy tensor. At early times the velocity gradient is largely longitudinal, which affects the stress-energy tensor by strengthening the transverse pressure, $T_{xx}=T_{yy}$, and decreasing the longitudinal pressure. In the Navier-Stokes equation,
\begin{equation}
\Delta T_{xx}=\frac{2\eta}{3\tau},~~\Delta T_{zz}=-\frac{4\eta}{3\tau},
\end{equation}
where $\eta$ is the shear viscosity and the velocity gradient for early times is $dv_z/dz=1/\tau$. After the first few fm/$c$, the transverse acceleration is determined by $T_{xx}$ and $T_{yy}$. As originally demonstrated in \cite{Romatschke:2007mq}, and shown here in the lower panel of Fig. \ref{fig:Rosl_everything}, the $R_{\rm out}/R_{\rm side}$ ratio can be lowered by $\sim 10\%$ with realistic shear viscosities. Analyses of elliptic flow have pointed to a small shear viscosity, perhaps approaching the KSS limit \cite{Kovtun:2004de}, $\eta_{\rm min}=s/4\pi$, where $s$ is the entropy density. The neglect of pre-thermalized flow in these calculations might have led to underestimates of the viscosity, but nonetheless, it is expected that $\eta$ is not much greater than the KSS bound. Below $T_c$, collisions are binary and the cascade prescription naturally accounts for viscous effects. Bulk viscosity is expected to be important near $T_c$ due to the inability of the system to maintain equilibrated chiral fields near $T_c$ \cite{Paech:2006st}. These expectations were also verified with lattice calculations \cite{Karsch:2007jc}. The impact of adding viscosity is shown in Fig. \ref{fig:Rosl_everything} (open triangles). Shear, which was set at twice the KSS bound, significantly affected the sources sizes, while changes to the bulk viscosity had little effect. Combined with the previously discussed effects, the calculation now approach the experimental data.

As a final improvement to the analysis, a more realistic scattering wave function was used in Eq. (\ref{eq:koonin}). The resulting correlation function was then fit to a Gaussian source using the approximate treatment applied in experimental treatments \cite{Bowler:1991vx,Sinyukov:1998fc}. Since sources are not purely Gaussian, and since the experimental fitting procedure only approximates the effects of Coulomb in the wave functions, the resulting radii differ slightly. The final results (filled circles in Fig. \ref{fig:Rosl_everything}) now match the data within $\approx$ 5\%, which is less than the normally quoted systematic experimental errors of 5-10\%. Thus, the HBT puzzle appears to have resulted from a conspiracy between several shortcomings in the original models, each of which led to an underestimate of the explosiveness of the collision, and each of which pushed $R_{\rm out}/R_{\rm side}$ in the same direction.

Other adjustments to the dynamics are known to affect femtoscopic radii. For the calculations presented here, the initial energy density profile was scaled to fit experimental charge multiplicity, while the shape of the profile was set by the wounded nucleon model \cite{Kolb:2000sd}. More compact profiles have been shown to increase the explosiveness and lower the $R_{\rm out}/R_{\rm side}$ ratio \cite{Broniowski:2008vp,Pratt:2008sz}. The model used here assumed boost-invariant accelerationless flow along the beam axis, similar to a simple Hubble expansion. This symmetry simplified and accelerated hydrodynamic calculations, but is only approximate, and is expected to cause errors of a few percent in HBT radii, most likely leading to a lowering of $R_{\rm long}$ by a few percent \cite{Pratt:2008jj}. Another effect not considered here concerns the interaction of the outgoing pions with a mean field from the remaining matter \cite{Cramer:2004ih}. Such effects only come into play after the final collisions of the pions, when densities are $\lesssim 0.1$ per fm$^3$ and fields are expected to be small. Mean field distortions are expected to affect radii only at the level of a few percent \cite{Pratt:2005bs}.

\begin{figure}
\centerline{\includegraphics[width=0.4\textwidth]{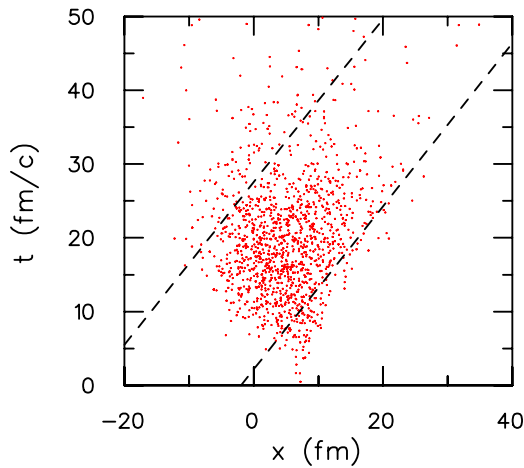}}
\caption{\label{fig:xyzt}(color online)
Final emission positions and times for particles with transverse momentum of 300 MeV/$c$ along the $x$ axis.  Emission has a positive correlation between position and time, though lags behind the slope of the velocity (illustrated by dashed lines). Due to the positive $x-t$ correlation, emissions of longer duration can still result in phase space clouds that are compact along the outward direction, with $R_{\rm out}/R_{\rm side}\approx 1$.}
\end{figure}

One of the puzzling aspects of HBT analyses came from fits with bast-wave models, which are parameteric pictures based on thermal emission from a collectively expanding source. Parameters include a breakup temperature $T$, an outer radius $R$, a breakup time $\tau$, a linearly rising transverse collective velocity with a maximum, $v_{\rm max}$, and an emission duration $\Delta\tau$. Blast wave fits are similar in quality to what was achieved here with a dynamic model. The surprising outcome of the fits was that radii were on the order of a dozen fm, with emission confined to within a few fm/$c$ of $\tau=10$ fm/$c$ \cite{Retiere:2003kf,Kisiel:2006is}. These parameters suggest an unphysically high breakup density. Figure \ref{fig:xyzt} shows the outward coordinate $x$ and the time $\tau$ at which emission occured for the final model in Fig. \ref{fig:Rosl_everything}, and are similar to what was seen in \cite{Lin:2002gc}. Points are shown only for particles emitted with momentum $p_x=300$ MeV/$c$. The dashed lines have a slope corresponding to the velocity $v_x=p_x/E$. Any two emissions at points connected by such a line contribute identically to the final phase space density, and thus identically to the femtoscopy. The positive correlation between $x$ and $t$ allows longer-lived emission to femtoscopically mimic more sudden emission at a much earlier time. This emphasizes the importance of using realistic dynamical models to compare to femtoscopic data and underscores the limits of parametric fits.

The calculations shown here were successful in matching experimental values of the mean transverse momentum for pions, kaons and protons. Elliptic flow, analyses of which are inherently precluded with the model used here due to an assumption of azimuthal symmetry, needs to be fit with the same parameters. More detailed aspects of femtoscopic data also need to be matched such as: radii with respect to the reaction plane \cite{Adams:2003ra}, correlations of other species, especially non-identical particles \cite{Kisiel:2004it,Adams:2003qa}, and non-Gaussian features of the source function \cite{Panitkin:2001qb}. Even if all these data are reproduced, it does not fully validate the model. That would require an ambitious statistical analysis of the set of model parameters and assumptions, similar to \cite{Habib:2007ca}. Although these goals require significant effort in the coming years, the current analysis has eliminated any puzzle about femtoscopy for the time being, as the experimental radii appear to be satisfactorily described within a rather standard theoretical picture of how RHIC collisions evolve. 

\begin{thebibliography}{38}
\expandafter\ifx\csname natexlab\endcsname\relax\def\natexlab#1{#1}\fi
\expandafter\ifx\csname bibnamefont\endcsname\relax
  \def\bibnamefont#1{#1}\fi
\expandafter\ifx\csname bibfnamefont\endcsname\relax
  \def\bibfnamefont#1{#1}\fi
\expandafter\ifx\csname citenamefont\endcsname\relax
  \def\citenamefont#1{#1}\fi
\expandafter\ifx\csname url\endcsname\relax
  \def\url#1{\texttt{#1}}\fi
\expandafter\ifx\csname urlprefix\endcsname\relax\def\urlprefix{URL }\fi
\providecommand{\bibinfo}[2]{#2}
\providecommand{\eprint}[2][]{\url{#2}}

\bibitem[{\citenamefont{Adcox et~al.}(2005)}]{Adcox:2004mh}
\bibinfo{author}{\bibfnamefont{K.}~\bibnamefont{Adcox}} \bibnamefont{et~al.}
  (\bibinfo{collaboration}{PHENIX}), \bibinfo{journal}{Nucl. Phys.}
  \textbf{\bibinfo{volume}{A757}}, \bibinfo{pages}{184} (\bibinfo{year}{2005}),
  \eprint{nucl-ex/0410003}.

\bibitem[{\citenamefont{Adams et~al.}(2005)}]{Adams:2005dq}
\bibinfo{author}{\bibfnamefont{J.}~\bibnamefont{Adams}} \bibnamefont{et~al.}
  (\bibinfo{collaboration}{STAR}), \bibinfo{journal}{Nucl. Phys.}
  \textbf{\bibinfo{volume}{A757}}, \bibinfo{pages}{102} (\bibinfo{year}{2005}),
  \eprint{nucl-ex/0501009}.

\bibitem[{\citenamefont{Heinz and Kolb}(2002)}]{Heinz:2002un}
\bibinfo{author}{\bibfnamefont{U.~W.} \bibnamefont{Heinz}} \bibnamefont{and}
  \bibinfo{author}{\bibfnamefont{P.~F.} \bibnamefont{Kolb}}
  (\bibinfo{year}{2002}), \eprint{hep-ph/0204061}.

\bibitem[{\citenamefont{Teaney et~al.}(2001)\citenamefont{Teaney, Lauret, and
  Shuryak}}]{Teaney:2001av}
\bibinfo{author}{\bibfnamefont{D.}~\bibnamefont{Teaney}},
  \bibinfo{author}{\bibfnamefont{J.}~\bibnamefont{Lauret}}, \bibnamefont{and}
  \bibinfo{author}{\bibfnamefont{E.~V.} \bibnamefont{Shuryak}}
  (\bibinfo{year}{2001}), \eprint{nucl-th/0110037}.

\bibitem[{\citenamefont{Soff et~al.}(2001)\citenamefont{Soff, Bass, and
  Dumitru}}]{Soff:2000eh}
\bibinfo{author}{\bibfnamefont{S.}~\bibnamefont{Soff}},
  \bibinfo{author}{\bibfnamefont{S.~A.} \bibnamefont{Bass}}, \bibnamefont{and}
  \bibinfo{author}{\bibfnamefont{A.}~\bibnamefont{Dumitru}},
  \bibinfo{journal}{Phys. Rev. Lett.} \textbf{\bibinfo{volume}{86}},
  \bibinfo{pages}{3981} (\bibinfo{year}{2001}), \eprint{nucl-th/0012085}.

\bibitem[{\citenamefont{Hirano and Tsuda}(2003)}]{Hirano:2002hv}
\bibinfo{author}{\bibfnamefont{T.}~\bibnamefont{Hirano}} \bibnamefont{and}
  \bibinfo{author}{\bibfnamefont{K.}~\bibnamefont{Tsuda}},
  \bibinfo{journal}{Nucl. Phys.} \textbf{\bibinfo{volume}{A715}},
  \bibinfo{pages}{821} (\bibinfo{year}{2003}), \eprint{nucl-th/0208068}.

\bibitem[{\citenamefont{Lisa et~al.}(2005)\citenamefont{Lisa, Pratt, Soltz, and
  Wiedemann}}]{Lisa:2005dd}
\bibinfo{author}{\bibfnamefont{M.~A.} \bibnamefont{Lisa}},
  \bibinfo{author}{\bibfnamefont{S.}~\bibnamefont{Pratt}},
  \bibinfo{author}{\bibfnamefont{R.}~\bibnamefont{Soltz}}, \bibnamefont{and}
  \bibinfo{author}{\bibfnamefont{U.}~\bibnamefont{Wiedemann}},
  \bibinfo{journal}{Ann. Rev. Nucl. Part. Sci.} \textbf{\bibinfo{volume}{55}},
  \bibinfo{pages}{357} (\bibinfo{year}{2005}), \eprint{nucl-ex/0505014}.

\bibitem[{\citenamefont{Hanbury~Brown and Twiss}(1956)}]{HanburyBrown:1956pf}
\bibinfo{author}{\bibfnamefont{R.}~\bibnamefont{Hanbury~Brown}}
  \bibnamefont{and} \bibinfo{author}{\bibfnamefont{R.~Q.} \bibnamefont{Twiss}},
  \bibinfo{journal}{Nature} \textbf{\bibinfo{volume}{178}},
  \bibinfo{pages}{1046} (\bibinfo{year}{1956}).

\bibitem[{\citenamefont{Petersen et~al.}(2008)\citenamefont{Petersen,
  Steinheimer, Li, Burau, and Bleicher}}]{Petersen:2008gy}
\bibinfo{author}{\bibfnamefont{H.}~\bibnamefont{Petersen}},
  \bibinfo{author}{\bibfnamefont{J.}~\bibnamefont{Steinheimer}},
  \bibinfo{author}{\bibfnamefont{Q.}~\bibnamefont{Li}},
  \bibinfo{author}{\bibfnamefont{G.}~\bibnamefont{Burau}}, \bibnamefont{and}
  \bibinfo{author}{\bibfnamefont{M.}~\bibnamefont{Bleicher}}
  (\bibinfo{year}{2008}), \eprint{0806.1805}.

\bibitem[{\citenamefont{Li et~al.}(2008)\citenamefont{Li, Bleicher, and
  Stocker}}]{Li:2008ge}
\bibinfo{author}{\bibfnamefont{Q.}~\bibnamefont{Li}},
  \bibinfo{author}{\bibfnamefont{M.}~\bibnamefont{Bleicher}}, \bibnamefont{and}
  \bibinfo{author}{\bibfnamefont{H.}~\bibnamefont{Stocker}},
  \bibinfo{journal}{Phys. Lett.} \textbf{\bibinfo{volume}{B663}},
  \bibinfo{pages}{395} (\bibinfo{year}{2008}), \eprint{0802.3618}.

\bibitem[{\citenamefont{Humanic}(2006)}]{Humanic:2006sk}
\bibinfo{author}{\bibfnamefont{T.~J.} \bibnamefont{Humanic}},
  \bibinfo{journal}{AIP Conf. Proc.} \textbf{\bibinfo{volume}{828}},
  \bibinfo{pages}{625} (\bibinfo{year}{2006}).

\bibitem[{\citenamefont{Lin et~al.}(2002)\citenamefont{Lin, Ko, and
  Pal}}]{Lin:2002gc}
\bibinfo{author}{\bibfnamefont{Z.-w.} \bibnamefont{Lin}},
  \bibinfo{author}{\bibfnamefont{C.~M.} \bibnamefont{Ko}}, \bibnamefont{and}
  \bibinfo{author}{\bibfnamefont{S.}~\bibnamefont{Pal}},
  \bibinfo{journal}{Phys. Rev. Lett.} \textbf{\bibinfo{volume}{89}},
  \bibinfo{pages}{152301} (\bibinfo{year}{2002}), \eprint{nucl-th/0204054}.

\bibitem[{\citenamefont{Pratt and Vredevoogd}(2008)}]{Pratt:2008sz}
\bibinfo{author}{\bibfnamefont{S.}~\bibnamefont{Pratt}} \bibnamefont{and}
  \bibinfo{author}{\bibfnamefont{J.}~\bibnamefont{Vredevoogd}}
  (\bibinfo{year}{2008}), \eprint{0809.0516}.

\bibitem[{\citenamefont{Koonin}(1977)}]{Koonin:1977fh}
\bibinfo{author}{\bibfnamefont{S.~E.} \bibnamefont{Koonin}},
  \bibinfo{journal}{Phys. Lett.} \textbf{\bibinfo{volume}{B70}},
  \bibinfo{pages}{43} (\bibinfo{year}{1977}).

\bibitem[{\citenamefont{Gyulassy et~al.}(2007)\citenamefont{Gyulassy, Sinyukov,
  Karpenko, and Nazarenko}}]{Gyulassy:2007zz}
\bibinfo{author}{\bibfnamefont{M.}~\bibnamefont{Gyulassy}},
  \bibinfo{author}{\bibfnamefont{Y.~M.} \bibnamefont{Sinyukov}},
  \bibinfo{author}{\bibfnamefont{I.}~\bibnamefont{Karpenko}}, \bibnamefont{and}
  \bibinfo{author}{\bibfnamefont{A.~V.} \bibnamefont{Nazarenko}},
  \bibinfo{journal}{Braz. J. Phys.} \textbf{\bibinfo{volume}{37}},
  \bibinfo{pages}{1031} (\bibinfo{year}{2007}).

\bibitem[{\citenamefont{Broniowski et~al.}(2008)\citenamefont{Broniowski,
  Chojnacki, Florkowski, and Kisiel}}]{Broniowski:2008vp}
\bibinfo{author}{\bibfnamefont{W.}~\bibnamefont{Broniowski}},
  \bibinfo{author}{\bibfnamefont{M.}~\bibnamefont{Chojnacki}},
  \bibinfo{author}{\bibfnamefont{W.}~\bibnamefont{Florkowski}},
  \bibnamefont{and} \bibinfo{author}{\bibfnamefont{A.}~\bibnamefont{Kisiel}},
  \bibinfo{journal}{Phys. Rev. Lett.} \textbf{\bibinfo{volume}{101}},
  \bibinfo{pages}{022301} (\bibinfo{year}{2008}), \eprint{0801.4361}.

\bibitem[{\citenamefont{Vredevoogd and Pratt}(2008)}]{Vredevoogd:2008id}
\bibinfo{author}{\bibfnamefont{J.}~\bibnamefont{Vredevoogd}} \bibnamefont{and}
  \bibinfo{author}{\bibfnamefont{S.}~\bibnamefont{Pratt}}
  (\bibinfo{year}{2008}), \eprint{0810.4325}.

\bibitem[{\citenamefont{Pratt}(1986)}]{Pratt:1986cc}
\bibinfo{author}{\bibfnamefont{S.}~\bibnamefont{Pratt}},
  \bibinfo{journal}{Phys. Rev.} \textbf{\bibinfo{volume}{D33}},
  \bibinfo{pages}{1314} (\bibinfo{year}{1986}).

\bibitem[{\citenamefont{Rischke and Gyulassy}(1996)}]{Rischke:1996em}
\bibinfo{author}{\bibfnamefont{D.~H.} \bibnamefont{Rischke}} \bibnamefont{and}
  \bibinfo{author}{\bibfnamefont{M.}~\bibnamefont{Gyulassy}},
  \bibinfo{journal}{Nucl. Phys.} \textbf{\bibinfo{volume}{A608}},
  \bibinfo{pages}{479} (\bibinfo{year}{1996}), \eprint{nucl-th/9606039}.

\bibitem[{\citenamefont{Hama et~al.}(2006)\citenamefont{Hama, Andrade, Grassi,
  {Socolowski Jr}, Kodama, Tavares, and Padula}}]{hama}
\bibinfo{author}{\bibfnamefont{Y.}~\bibnamefont{Hama}},
  \bibinfo{author}{\bibfnamefont{R.~P.~G.} \bibnamefont{Andrade}},
  \bibinfo{author}{\bibfnamefont{F.}~\bibnamefont{Grassi}},
  \bibinfo{author}{\bibfnamefont{O.}~\bibnamefont{{Socolowski Jr}}},
  \bibinfo{author}{\bibfnamefont{T.}~\bibnamefont{Kodama}},
  \bibinfo{author}{\bibfnamefont{B.}~\bibnamefont{Tavares}}, \bibnamefont{and}
  \bibinfo{author}{\bibfnamefont{S.~S.} \bibnamefont{Padula}},
  \bibinfo{journal}{Nuclear Physics A} \textbf{\bibinfo{volume}{774}},
  \bibinfo{pages}{169} (\bibinfo{year}{2006}),
  \urlprefix\url{http://www.citebase.org/abstract?id=oai:arXiv.org:hep-ph/0510%
096}.

\bibitem[{\citenamefont{Pratt}(2008)}]{Pratt:2008sc}
\bibinfo{author}{\bibfnamefont{S.}~\bibnamefont{Pratt}} (\bibinfo{year}{2008}),
  \eprint{0809.0089}.

\bibitem[{\citenamefont{Romatschke and Romatschke}(2007)}]{Romatschke:2007mq}
\bibinfo{author}{\bibfnamefont{P.}~\bibnamefont{Romatschke}} \bibnamefont{and}
  \bibinfo{author}{\bibfnamefont{U.}~\bibnamefont{Romatschke}},
  \bibinfo{journal}{Phys. Rev. Lett.} \textbf{\bibinfo{volume}{99}},
  \bibinfo{pages}{172301} (\bibinfo{year}{2007}), \eprint{0706.1522}.

\bibitem[{\citenamefont{Kovtun et~al.}(2005)\citenamefont{Kovtun, Son, and
  Starinets}}]{Kovtun:2004de}
\bibinfo{author}{\bibfnamefont{P.}~\bibnamefont{Kovtun}},
  \bibinfo{author}{\bibfnamefont{D.~T.} \bibnamefont{Son}}, \bibnamefont{and}
  \bibinfo{author}{\bibfnamefont{A.~O.} \bibnamefont{Starinets}},
  \bibinfo{journal}{Phys. Rev. Lett.} \textbf{\bibinfo{volume}{94}},
  \bibinfo{pages}{111601} (\bibinfo{year}{2005}), \eprint{hep-th/0405231}.

\bibitem[{\citenamefont{Paech and Pratt}(2006)}]{Paech:2006st}
\bibinfo{author}{\bibfnamefont{K.}~\bibnamefont{Paech}} \bibnamefont{and}
  \bibinfo{author}{\bibfnamefont{S.}~\bibnamefont{Pratt}},
  \bibinfo{journal}{Phys. Rev.} \textbf{\bibinfo{volume}{C74}},
  \bibinfo{pages}{014901} (\bibinfo{year}{2006}), \eprint{nucl-th/0604008}.

\bibitem[{\citenamefont{Karsch et~al.}(2008)\citenamefont{Karsch, Kharzeev, and
  Tuchin}}]{Karsch:2007jc}
\bibinfo{author}{\bibfnamefont{F.}~\bibnamefont{Karsch}},
  \bibinfo{author}{\bibfnamefont{D.}~\bibnamefont{Kharzeev}}, \bibnamefont{and}
  \bibinfo{author}{\bibfnamefont{K.}~\bibnamefont{Tuchin}},
  \bibinfo{journal}{Phys. Lett.} \textbf{\bibinfo{volume}{B663}},
  \bibinfo{pages}{217} (\bibinfo{year}{2008}), \eprint{0711.0914}.

\bibitem[{\citenamefont{Bowler}(1991)}]{Bowler:1991vx}
\bibinfo{author}{\bibfnamefont{M.~G.} \bibnamefont{Bowler}},
  \bibinfo{journal}{Phys. Lett.} \textbf{\bibinfo{volume}{B270}},
  \bibinfo{pages}{69} (\bibinfo{year}{1991}).

\bibitem[{\citenamefont{Sinyukov et~al.}(1998)\citenamefont{Sinyukov, Lednicky,
  Akkelin, Pluta, and Erazmus}}]{Sinyukov:1998fc}
\bibinfo{author}{\bibfnamefont{Y.}~\bibnamefont{Sinyukov}},
  \bibinfo{author}{\bibfnamefont{R.}~\bibnamefont{Lednicky}},
  \bibinfo{author}{\bibfnamefont{S.~V.} \bibnamefont{Akkelin}},
  \bibinfo{author}{\bibfnamefont{J.}~\bibnamefont{Pluta}}, \bibnamefont{and}
  \bibinfo{author}{\bibfnamefont{B.}~\bibnamefont{Erazmus}},
  \bibinfo{journal}{Phys. Lett.} \textbf{\bibinfo{volume}{B432}},
  \bibinfo{pages}{248} (\bibinfo{year}{1998}).

\bibitem[{\citenamefont{Kolb et~al.}(2000)\citenamefont{Kolb, Sollfrank, and
  Heinz}}]{Kolb:2000sd}
\bibinfo{author}{\bibfnamefont{P.~F.} \bibnamefont{Kolb}},
  \bibinfo{author}{\bibfnamefont{J.}~\bibnamefont{Sollfrank}},
  \bibnamefont{and} \bibinfo{author}{\bibfnamefont{U.~W.} \bibnamefont{Heinz}},
  \bibinfo{journal}{Phys. Rev.} \textbf{\bibinfo{volume}{C62}},
  \bibinfo{pages}{054909} (\bibinfo{year}{2000}), \eprint{hep-ph/0006129}.

\bibitem[{\citenamefont{Pratt}(2007)}]{Pratt:2008jj}
\bibinfo{author}{\bibfnamefont{S.}~\bibnamefont{Pratt}},
  \bibinfo{journal}{Phys. Rev.} \textbf{\bibinfo{volume}{C75}},
  \bibinfo{pages}{024907} (\bibinfo{year}{2007}), \eprint{nucl-th/0612010}.

\bibitem[{\citenamefont{Cramer et~al.}(2005)\citenamefont{Cramer, Miller, Wu,
  and Yoon}}]{Cramer:2004ih}
\bibinfo{author}{\bibfnamefont{J.~G.} \bibnamefont{Cramer}},
  \bibinfo{author}{\bibfnamefont{G.~A.} \bibnamefont{Miller}},
  \bibinfo{author}{\bibfnamefont{J.~M.~S.} \bibnamefont{Wu}}, \bibnamefont{and}
  \bibinfo{author}{\bibfnamefont{J.-H.} \bibnamefont{Yoon}},
  \bibinfo{journal}{Phys. Rev. Lett.} \textbf{\bibinfo{volume}{94}},
  \bibinfo{pages}{102302} (\bibinfo{year}{2005}), \eprint{nucl-th/0411031}.

\bibitem[{\citenamefont{Pratt}(2006)}]{Pratt:2005bs}
\bibinfo{author}{\bibfnamefont{S.}~\bibnamefont{Pratt}}, \bibinfo{journal}{AIP
  Conf. Proc.} \textbf{\bibinfo{volume}{828}}, \bibinfo{pages}{213}
  (\bibinfo{year}{2006}), \eprint{nucl-th/0511009}.

\bibitem[{\citenamefont{Retiere and Lisa}(2004)}]{Retiere:2003kf}
\bibinfo{author}{\bibfnamefont{F.}~\bibnamefont{Retiere}} \bibnamefont{and}
  \bibinfo{author}{\bibfnamefont{M.~A.} \bibnamefont{Lisa}},
  \bibinfo{journal}{Phys. Rev.} \textbf{\bibinfo{volume}{C70}},
  \bibinfo{pages}{044907} (\bibinfo{year}{2004}), \eprint{nucl-th/0312024}.

\bibitem[{\citenamefont{Kisiel et~al.}(2006)\citenamefont{Kisiel, Florkowski,
  and Broniowski}}]{Kisiel:2006is}
\bibinfo{author}{\bibfnamefont{A.}~\bibnamefont{Kisiel}},
  \bibinfo{author}{\bibfnamefont{W.}~\bibnamefont{Florkowski}},
  \bibnamefont{and}
  \bibinfo{author}{\bibfnamefont{W.}~\bibnamefont{Broniowski}},
  \bibinfo{journal}{Phys. Rev.} \textbf{\bibinfo{volume}{C73}},
  \bibinfo{pages}{064902} (\bibinfo{year}{2006}), \eprint{nucl-th/0602039}.

\bibitem[{\citenamefont{Adams et~al.}(2004)}]{Adams:2003ra}
\bibinfo{author}{\bibfnamefont{J.}~\bibnamefont{Adams}} \bibnamefont{et~al.}
  (\bibinfo{collaboration}{STAR}), \bibinfo{journal}{Phys. Rev. Lett.}
  \textbf{\bibinfo{volume}{93}}, \bibinfo{pages}{012301}
  (\bibinfo{year}{2004}), \eprint{nucl-ex/0312009}.

\bibitem[{\citenamefont{Kisiel}(2004)}]{Kisiel:2004it}
\bibinfo{author}{\bibfnamefont{A.}~\bibnamefont{Kisiel}}
  (\bibinfo{collaboration}{STAR}), \bibinfo{journal}{J. Phys.}
  \textbf{\bibinfo{volume}{G30}}, \bibinfo{pages}{S1059}
  (\bibinfo{year}{2004}), \eprint{nucl-ex/0403042}.

\bibitem[{\citenamefont{Adams et~al.}(2003)}]{Adams:2003qa}
\bibinfo{author}{\bibfnamefont{J.}~\bibnamefont{Adams}} \bibnamefont{et~al.}
  (\bibinfo{collaboration}{STAR}), \bibinfo{journal}{Phys. Rev. Lett.}
  \textbf{\bibinfo{volume}{91}}, \bibinfo{pages}{262302}
  (\bibinfo{year}{2003}), \eprint{nucl-ex/0307025}.

\bibitem[{\citenamefont{Panitkin et~al.}(2001)}]{Panitkin:2001qb}
\bibinfo{author}{\bibfnamefont{S.~Y.} \bibnamefont{Panitkin}}
  \bibnamefont{et~al.} (\bibinfo{collaboration}{E895}), \bibinfo{journal}{Phys.
  Rev. Lett.} \textbf{\bibinfo{volume}{87}}, \bibinfo{pages}{112304}
  (\bibinfo{year}{2001}), \eprint{nucl-ex/0103011}.

\bibitem[{\citenamefont{Habib et~al.}(2007)\citenamefont{Habib, Heitmann,
  Higdon, Nakhleh, and Williams}}]{Habib:2007ca}
\bibinfo{author}{\bibfnamefont{S.}~\bibnamefont{Habib}},
  \bibinfo{author}{\bibfnamefont{K.}~\bibnamefont{Heitmann}},
  \bibinfo{author}{\bibfnamefont{D.}~\bibnamefont{Higdon}},
  \bibinfo{author}{\bibfnamefont{C.}~\bibnamefont{Nakhleh}}, \bibnamefont{and}
  \bibinfo{author}{\bibfnamefont{B.}~\bibnamefont{Williams}},
  \bibinfo{journal}{Phys. Rev.} \textbf{\bibinfo{volume}{D76}},
  \bibinfo{pages}{083503} (\bibinfo{year}{2007}), \eprint{astro-ph/0702348}.

\end{thebibliography}
\end{document}